\documentclass
[notitlepage,12pt,groupedaddress,secnumarabic,prd,showpacs,showkeys,onecolumn,nofootinbib]{revtex4}%
\usepackage{amsfonts}
\usepackage{amsmath}
\usepackage{amssymb}
\usepackage{graphicx}
\usepackage{appendix}
\usepackage{color}%
\begin{document}
\title{Radially symmetric scalar solitons}
\author{J.R. Morris}
\affiliation{Physics Department, Indiana University Northwest, 3400 Broadway, Gary, IN
46408, USA}
\email{jmorris@iun.edu}

\begin{abstract}
A class of noncanonical effective potentials is introduced allowing stable,
radially symmetric, solutions to first order Bogomol'nyi equations for a real
scalar field in a fixed spacetime background. This class of effective
potentials generalizes those found previously by Bazeia, Menezes, and Menezes
[Phys.Rev.Lett. 91 (2003) 241601] for radially symmetric defects in a flat
spacetime. Use is made of the \textquotedblleft on-shell
method\textquotedblright\ introduced by Atmaja and Ramadhan [Phys.Rev.D 90
(2014) 10, 105009] of reducing the second order equation of motion to a first
order one, along with a constraint equation. This method and class of
potentials admits radially symmetric, stable solutions for four dimensional
static, radially symmetric spacetimes. Stability against radial fluctuations
is established with a modified version of Derrick's theorem, along with
demonstrating that the radial stress vanishes. Several examples of scalar
field configurations are given.

\end{abstract}

\maketitle

\section{Introduction}

\ \ A class of noncanonical scalar field potentials of the form $U(r,\phi
)=r^{-N}P(\phi)$, where $r$ is a radial variable, was introduced by Bazeia,
Menezes, and Menezes \cite{Bazeia PRL03} for scalar field theories in $D$
space dimensions. It was shown that for certain constraints for $D$ and $N$
that radially stable scalar field configurations can exist, evading Derrick's
theorem \cite{Derrick}. The $r^{-N}$ factor in $U(r,\phi)$ can emerge from a
more fundamental theory which gives rise to an effective scalar field model.
Potentials of this form have been physically motivated and considered in
various contexts \cite{Bazeia PRD18},\cite{Bazeia PRR19},\cite{Andrade
PRD19},\cite{Casana PRD20},\cite{Bazeia ppt20}. They give rise to field
theoretic models with interesting properties, and are of mathematical
interest, as well.

\bigskip

\ \ Attention has also been focused upon the possibility of evading Derrick's
theorem within the context of replacing a flat spacetime by curved spacetimes.
(See, for example, \cite{Gonzalez RMF01} - \cite{Perivol PRD20}.) However,
most of this work has been applied to systems with canonical scalar field
potentials, which have no explicit dependence upon a radial variable, with an
emphasis upon the effects of spacetime curvature on the stability of solitonic systems.

\bigskip

\ \ Presently, another class of potentials of the more general form
$V(r,\phi)=F(r)P(\phi)$ is introduced in order to solve first order
Bogomol'nyi equations for a scalar field $\phi(r)$ in a fixed static, radially
symmetric four dimensional spacetime background. The nontopological solitonic
solutions depend upon the assumed form of $P(\phi)$ and the form of the
spacetime metric. These solutions are also found to minimize the energy and to
be radially stable. \ The radial stability can be established with a modified
version of Derrick's theorem \cite{Derrick}, along with showing that the
radial stress vanishes. The first order Bogomol'nyi equations can be solved
using the \textquotedblleft on-shell\textquotedblright\ type of method
introduced by Atmaja and Ramadhan \cite{Atmaja PRD14} whereby a term is added
and subtracted from the second order Euler-Lagrange equations of motion,
thereby allowing a split of the second order differential equation (DE) into
one first order Bogomol'nyi DE plus a constraint equation. (See also,
\cite{Adam JHEP16},\cite{Atmaja JHEP16}.) The solution to these equations
automatically satisfies the second order equation of motion. Here, this
procedure is adapted to a single static, radially symmetric scalar field
$\phi(r)$ for a particular type of potential $V(r,\phi)=F(r)P(\phi)$ whose
function $F(r)$ depends upon the spacetime metric $g_{\mu\nu}$. Furthermore,
this solution $\phi(r)$ provides a lower bound on the energy, and the scalar
configuration described by $\phi(r)$ is shown to be stable against radial
collapse or expansion.

\bigskip

\ \ The possibility of scalarization of gravitational sources, such as neutron
stars, was introduced by Damour and Esposito-Farese \cite{Damour PRL93} in the
context of scalar-tensor theory expressed in an Einstein frame. There, the
response of a scalar field near a gravitational source was due to strong field
effects associated with high curvature. Although a spontaneous scalarization
can occur due to gravitational effects, the concept of scalarization can be
extended to situations where other fields are involved. An example is provided
by Maxwell-scalar theory, where a real valued scalar field couples
nonminimally to the Maxwell field via a coupling function, say $\varepsilon
(\phi)$, through an interaction term $-\frac{1}{4}\varepsilon(\phi)F_{\mu\nu
}F^{\mu\nu}$. A scalar field may form around a compact object, even in a flat
Minkowski spacetime (see, e,g, \cite{Bazeia ppt20} and \cite{Herdeiro PRD21},
and references therein). Models involving the formation of a scalar cloud
around some source in a fixed spacetime background can serve as toy models of
more realistic processes where back reactions upon the spacetime can be taken
into account. Here, we examine the case of a scalar field responding to
another, unspecified, field through an effective scalar potential $V(r,\phi)$.
In particular, we consider the radially stable solutions of a first order
Bogomol'nyi equation.

\bigskip

\ \ Examples of how an effective scalar potential can arise from interactions
with other fields are provided in Section 2,\ and a BPS ansatz for obtaining
Bogomol'nyi equations and solutions is presented in Section 3. The radial
stability of the solutions is discussed in Section 4. Several examples of
application are provided in Section 5. Section 6 concludes with a brief
summary. Details regarding stability arguments are relegated to an appendix.

\section{An effective potential}

\ \ A potential of the form $V(r,\phi)=F(r)P(\phi)=\frac{1}{f^{2}h}P(\phi)$
that is considered here can arise naturally as an effective scalar potential
for a real scalar field $\phi$ that interacts with another field. Two such
examples are provided here.

\subsection{Interacting scalars}

\ \ Consider a model of two interacting real scalar fields, $\phi$ and $\chi$,
described by%
\begin{equation}
\mathcal{L}=\frac{1}{2}\partial_{\mu}\phi\partial^{\mu}\phi+\frac{1}{2}%
K(\phi)\partial_{\mu}\chi\partial^{\mu}\chi\label{e1}%
\end{equation}

The equations of motion are given by
\begin{subequations}
\label{e2}%
\begin{align}
\nabla_{\mu}\nabla^{\mu}\phi-\frac{1}{2}(\partial_{\phi}K)(\partial_{\mu}%
\chi\partial^{\mu}\chi)  &  =0\label{e2a}\\
\frac{1}{2}\nabla_{\mu}[K(\phi)\partial^{\mu}\chi]  &  =0 \label{e2b}%
\end{align}

Assume now a radially symmetric (i.e., spherical or cylindrical symmetry)
ansatz for time independent fields in a static, radially symmetric, four
dimensional spacetime, $\phi=\phi(r)$ and $\chi=\chi(r)$. Also, denote
$g=|\det g_{\mu\nu}|$ and let the radial part of $\sqrt{g}$ be designated by
$f(r)$ and define $|g^{rr}|=h(r)$ where $r$ is the radial variable. In this
case the equations of motion (\ref{e2}) reduce to
\end{subequations}
\begin{subequations}
\label{e3}%
\begin{align}
-\frac{1}{f}\partial_{r}(fh\partial_{r}\phi)+\frac{1}{2}(\partial_{\phi
}K)[h(\partial_{r}\chi)^{2}]  &  =0\label{e3a}\\
\partial_{r}[fhK(\phi)\partial_{r}\chi]  &  =0 \label{e3b}%
\end{align}

From (\ref{e3b}) it follows that%
\end{subequations}
\begin{equation}
\partial_{r}\chi=\frac{C}{fhK(\phi)} \label{e4}%
\end{equation}

where $C$ is a constant. From (\ref{e4}) we have%
\begin{equation}
\frac{1}{2}K(\phi)\partial_{\mu}\chi\partial^{\mu}\chi=\frac{1}{2}%
K(\phi)(\partial_{r}\chi\partial^{r}\chi)=-\frac{1}{2}Kh(\partial_{r}\chi
)^{2}=-\frac{1}{2f^{2}h}\frac{C^{2}}{K(\phi)} \label{e5}%
\end{equation}

Using (\ref{e5}), (\ref{e1}) yields an effective Lagrangian for the field
$\phi$, $\mathcal{L}\rightarrow\frac{1}{2}\partial_{\mu}\phi\partial^{\mu}%
\phi-V(r,\phi)$, where the effective potential is%
\begin{equation}
V(r,\phi)=-\frac{1}{2}K(\phi)\partial_{\mu}\chi\partial^{\mu}\chi=\frac
{1}{2f^{2}h}\frac{C^{2}}{K(\phi)}=F(r)P(\phi) \label{e6}%
\end{equation}

where $F(r)=1/(f^{2}h)$ and $P(\phi)=\frac{1}{2}C^{2}K^{-1}(\phi)$. The
equation of motion for $\phi$ can now be written as%
\begin{equation}
\square\phi+\partial_{\phi}V(r,\phi)=0 \label{e7}%
\end{equation}

\subsection{Maxwell-scalar theory}

\ \ A second example is provided by a Maxwell-scalar model involving a real,
massless scalar field $\phi$ coupled nonminimally to an abelian Maxwell field
$F_{\mu\nu}$. The Lagrangian is
\begin{equation}
\mathcal{L}=\frac{1}{2}\partial_{\mu}\phi\partial^{\mu}\phi-\frac{1}%
{4}\varepsilon(r,\phi)F_{\mu\nu}F^{\mu\nu}-J^{\nu}A_{\nu} \label{e8}%
\end{equation}

where $\varepsilon(r,\phi)$ is a nonminimal coupling function which may
display a tachyonic instability that can depend upon the radial distance $r$
from the coordinate origin. We want to consider the spatial region exterior to
the source of charge $Q$, i.e., the region of space where $J^{\nu}%
\rightarrow0$. The equations of motion that follow from (\ref{e8}) are%
\begin{equation}%
\begin{array}
[c]{cc}%
\nabla_{\mu}\nabla^{\mu}\phi+\frac{1}{4}(\partial_{\phi}\varepsilon)F_{\mu\nu
}F^{\mu\nu}+\partial_{\phi}J^{\nu}A_{\nu}=0\smallskip & \\
\nabla_{\mu}(\varepsilon F^{\mu\nu})=J^{\nu} &
\end{array}
\label{e9}%
\end{equation}

(A metric with signature $(+,-,-,-)$ is assumed. See Eq.(\ref{1}) below.)
Using $F_{\mu\nu}F^{\mu\nu}=-2(\mathbf{E}^{2}-\mathbf{B}^{2})$ and setting
$J^{\nu}=0$ and $\mathbf{B}=0$, these reduce to%
\begin{equation}%
\begin{array}
[c]{cc}%
\nabla_{\mu}\nabla^{\mu}\phi-\frac{1}{2}(\partial_{\phi}\varepsilon
)\mathbf{E}^{2}=0\smallskip & \\
\nabla_{\mu}(\varepsilon F^{\mu\nu})=0 &
\end{array}
\label{e10}%
\end{equation}

Assuming radial symmetry, the Maxwell equation reduces to $\nabla
_{r}(\varepsilon F^{r0})=\frac{1}{\sqrt{g}}\partial_{r}(\sqrt{g}\varepsilon
F^{r0})=\frac{1}{f}\partial_{r}(f\varepsilon F^{r0})=0$ which is solved by%
\begin{equation}
F^{r0}(r,\phi)=\frac{Q}{f(r)\varepsilon(r,\phi)},\ \ \ F_{r0}=g_{rr}%
g_{00}F^{r0}=-\frac{g_{00}Q}{f(r)h(r)\varepsilon(r,\phi)} \label{e11}%
\end{equation}

where $g_{rr}=1/g^{rr}$ and we define a \textquotedblleft rationalized
charge\textquotedblright\ $Q=Q_{0}/4\pi$, (or \textquotedblleft rationalized
linear charge density\textquotedblright\ $Q=Q_{0}/2\pi$) with $Q_{0}$
representing the actual charge/charge density, and%
\begin{equation}
\mathbf{E}^{2}=-F_{r0}F^{r0}=\frac{g_{00}|g_{rr}|Q^{2}}{f^{2}\varepsilon^{2}%
}=\frac{g_{00}Q^{2}}{f^{2}h\varepsilon^{2}} \label{e12}%
\end{equation}

\ \ An effective Lagrangian for the scalar field $\phi$ is%
\begin{equation}
\mathcal{L}=\frac{1}{2}\partial_{\mu}\phi\partial^{\mu}\phi+\frac{1}%
{2}\varepsilon(r,\phi)\mathbf{E}^{2}=\frac{1}{2}\partial_{\mu}\phi
\partial^{\mu}\phi+\ \frac{g_{00}Q^{2}}{2f^{2}h}\varepsilon^{-1}\ (r,\phi)
\label{e13}%
\end{equation}

Now define an effective potential%
\begin{equation}
V(r,\phi)=\frac{1}{2}\varepsilon\mathbf{E}^{2}=\frac{g_{00}Q^{2}}{2f^{2}%
h}\varepsilon^{-1}(r,\phi) \label{e14}%
\end{equation}

An implementation of a BPS ansatz (presented below) will allow the coupling
function to be expressed as $g_{00}Q^{2}\varepsilon^{-1}(r,\phi)=X^{2}(\phi)$.
Using the expression (\ref{e12}) for $\mathbf{E}^{2}(r,\phi)\propto
\varepsilon^{-2}(r,\phi)$, we see that for the ansatz solutions $\phi(r)$ we
have, by (\ref{e14}), $V(r,\phi)=\frac{1}{2}\varepsilon\mathbf{E}%
^{2}\rightarrow\frac{1}{2f^{2}h}X^{2}$, and $-\frac{1}{2}(\partial_{\phi
}\varepsilon)\mathbf{E}^{2}=+\partial_{\phi}(\frac{1}{2}\varepsilon
\mathbf{E}^{2})$, so that an effective scalar potential appearing in the
equation of motion for $\phi$ has the form $V(r,\phi)=\frac{1}{2}%
\varepsilon(r,\phi)\mathbf{E}^{2}(r,\phi)\rightarrow\frac{1}{2f^{2}h}%
X^{2}(\phi)$. In this case the scalar field equation of motion takes its
standard form $\square\phi+\partial_{\phi}V(r,\phi)=0$.

\bigskip

\ \ To summarize, a scalar field $\phi$ can interact with other fields (e.g.,
scalars and gauge fields \cite{Atmaja PRD14}-\cite{Atmaja JHEP16}) with an
effective scalar potential of the form $V(r,\phi)=\frac{1}{f^{2}h}P(\phi)$.
Scalar models with such an effective potential will be seen to have radially
stable time independent solutions that obey a first order Bogomol'nyi equation.

\section{BPS ansatz}

\ \ The spacetime geometries considered here are assumed to be fixed, i.e.,
back reactions of the scalar field upon the metric are ignored\footnote{The
background metric is fixed and is therefore not required to solve any
particular equation of motion.}. The scalar field is assumed to be minimally
coupled to the gravitational sector, that is, the action is written in an
Einstein frame. We consider 4D metrics with radial symmetry (spherical or
cylindrical) of the form%
\begin{equation}
\left\{
\begin{array}
[c]{ll}%
ds^{2}=A(r)dt^{2}-B(r)dr^{2}-R^{2}(r)(d\theta^{2}+\sin^{2}\theta d\varphi
^{2}), & \text{(spherical symmetry) }\medskip\\
ds^{2}=A(r)dt^{2}-B(r)dr^{2}-\rho^{2}(r)d\varphi^{2}-\zeta^{2}(r)dz^{2}, &
\text{(cylindrical symmetry)}%
\end{array}
\right.  \label{1}%
\end{equation}

in which case we have $\sqrt{g}=\sqrt{AB}R^{2}\sin\theta$ for spherical
symmetry and $\sqrt{g}=\sqrt{AB}\rho\zeta$ for cylindrical symmetry
\cite{Trenda EJP}, where $g=|\det g_{\mu\nu}|$. (The functions $A(r)$ and
$B(r)$ for the cylindrical case are generally different from those for the
spherical case.) We represent the radial part of $\sqrt{g}$ by%
\begin{equation}
f(r)=\sqrt{A(r)B(r)C(r)} \label{14a}%
\end{equation}

where $\sqrt{C(r)}$ is given by%
\begin{equation}
\sqrt{C(r)}=\left\{
\begin{array}
[c]{ll}%
R^{2}(r), & \text{(spherical symmetry)\medskip}\\
\rho(r)\zeta(r), & \text{(cylindrical symmetry)}%
\end{array}
\right.  \label{14b}%
\end{equation}

\bigskip

\ \ The Lagrangian for the real scalar field $\phi$ is given by%
\begin{equation}
\mathcal{L}=\frac{1}{2}\partial^{\mu}\phi\partial_{\mu}\phi-V(r,\phi
),\ \ \ V(r,\phi)=F(r)P(\phi) \label{2}%
\end{equation}

where the noncanonical potential $V(r,\phi)=F(r)P(\phi)$ depends not only upon
the scalar field $\phi$, but also has an explicit dependence upon the radial
coordinate $r$, as in Ref \cite{Bazeia PRL03}. We consider static, radially
symmetric solutions $\phi(r)$, for which the Lagrangian can be written as%
\begin{equation}
\mathcal{L}=\frac{1}{2}g^{rr}(r)(\partial_{r}\phi)^{2}-F(r)P(\phi) \label{3}%
\end{equation}

where $g^{rr}(r)=1/g_{rr}(r)=-1/B(r)$ and $\partial_{r}=\partial/\partial r$.

\bigskip

\ \ The equation of motion following from (\ref{2}) is given by $\nabla_{\mu
}\nabla^{\mu}\phi+\partial_{\phi}V(r,\phi)=0,$ or
\begin{equation}
\frac{1}{\sqrt{g}}\partial_{r}(\sqrt{g}g^{rr}\partial_{r}\phi)+\partial_{\phi
}V=0 \label{4}%
\end{equation}
for $\phi=\phi(r)$, with $\partial_{\phi}V=\partial V/\partial\phi$. We can
also define $h(r)\equiv|g^{rr}(r)|=1/B(r)$, or $g^{rr}=-h$ for our metric
signature. The equation of motion (\ref{4}) then reduces to%
\begin{equation}
\partial_{r}[f(r)h(r)\partial_{r}\phi]=f(r)\partial_{\phi}V(r,\phi) \label{5}%
\end{equation}

\ \ We now use the method of Atmaja and Ramadhan \cite{Atmaja PRD14} to
generate a first order Bogomol'nyi equation by subtracting a term
$\partial_{r}X(\phi)$ from both sides of (\ref{5}):%
\begin{equation}
\partial_{r}[f(r)h(r)\partial_{r}\phi-X(\phi)]=f(r)\partial_{\phi}%
V(r,\phi)-\partial_{r}X(\phi) \label{6}%
\end{equation}

where the function $X=X(\phi)$ and $\partial_{r}X(\phi)=\partial_{\phi}%
X(\phi)\partial_{r}\phi$. The Euler-Lagrange equation of motion, i.e.,
equation (\ref{6}) is then solved by solutions to the set of equations%
\begin{equation}
f(r)h(r)\partial_{r}\phi=X(\phi),\ \ \ f(r)\partial_{\phi}V(r,\phi
)=\partial_{\phi}X(\phi)\partial_{r}\phi\label{7}%
\end{equation}

The first equation is the first order Bogomol'nyi equation, and the second
equation gives the form of the potential $V$ in terms of $X(\phi)$ and $r$,
since
\begin{equation}
\partial_{\phi}V=\partial_{\phi}X\cdot\frac{1}{f^{2}h}X=\frac{1}{2f^{2}%
h}\partial_{\phi}X^{2} \label{8}%
\end{equation}

Integrating gives a potential $V=(2f^{2}h)^{-1}(X^{2}+c)$. Setting the
constant $c=0$ and requiring that the function $X(\phi)$ is chosen so that $V$
is everywhere finite for a finite energy solution, then yields%
\begin{equation}
V(r,\phi)=\frac{1}{2f^{2}(r)h(r)}X^{2}(\phi) \label{9}%
\end{equation}

where $1/h=|g_{rr}|=B$. The second order equation of motion is then reduced to
a first order one, along with a constraint on the form of $V(r,\phi)$:%
\begin{equation}
\partial_{r}\phi(r)=\frac{1}{fh}X(\phi),\ \ \ \ \ V(r,\phi)=\frac{1}{2f^{2}%
h}X^{2}(\phi)=F(r)P(\phi) \label{10}%
\end{equation}

We can identify $F(r)=(f^{2}h)^{-1}$ and $P(\phi)=\frac{1}{2}X^{2}(\phi)$.
Upon choosing a suitable form for $X(\phi)$ that keeps the energy $E$ (or,
energy per unit length for cylindrical symmetry) of the scalar field
configuration finite, the Bogomol'nyi equation is solved by
\begin{subequations}
\label{11}%
\begin{align}
\int\frac{d\phi}{X(\phi)}  &  =\int\frac{dr}{f(r)h(r)},\label{11a}\\
V(r,\phi)  &  =\frac{1}{2f^{2}(r)h(r)}X^{2}(\phi) \label{11b}%
\end{align}

We note that for $h=1$ and $f=r^{N}$ this coincides with the form of the
potential introduced in Ref. \cite{Bazeia PRL03} for flat spacetimes, with the
identification $X(\phi)=W_{\phi}(\phi)=\partial_{\phi}W$, where $W(\phi)$ is a superpotential.

\section{Energy and stability}

\ \ \ \ \ \ The component of the stress-energy tensor associated with the
energy density of the static scalar field $\phi(r)$ is $T_{0}^{0}%
=-\mathcal{L}=-[\frac{1}{2}g^{rr}(\partial_{r}\phi)^{2}-V]$, which we will
also label as $\mathcal{H}$. From (\ref{1}) we have $g^{rr}=-h(r)=-B^{-1}(r)$
so that $\mathcal{H}=\frac{1}{2}h(\partial_{r}\phi)^{2}+V$. From (\ref{10}) we
have gradient and potential contributions $\mathcal{H}_{g}=\frac{1}%
{2}h(\partial_{r}\phi)^{2}$ and $\mathcal{H}_{p}=V(r,\phi)=\frac{1}{2f^{2}%
h}X^{2}(\phi)=\frac{1}{f^{2}h}P(\phi)$. For the ansatz solutions (\ref{10})
the gradient and potential parts are connected by $X(\phi)$ and contribute
equally, but for an arbitrary solution to the second order equation of motion
we consider the gradient and potential parts separately in applying an
approach to analyze stability. Stability for the ansatz solutions is then
demonstrated by connecting the gradient and potential pieces.

\bigskip

\ \ The energy of the scalar field is $E=\int T_{0}^{0}\sqrt{g}d^{3}x$, so
that upon removing the integrations over the nonradial coordinates (see
Appendix) we have an energy parameter $\mathcal{E}$, given by%
\end{subequations}
\begin{equation}
\mathcal{E}=\int T_{0}^{0}f(r)dr=\int\mathcal{H}f(r)dr \label{12}%
\end{equation}

For a stable solution $\phi(r)$ we require $\mathcal{E}$ to be finite, and to
represent a stable minimum of the energy. In order to determine whether a
static, radially dependent solution $\phi(r)$ represents a stable minimum of
the action and energy, we follow the line of reasoning used in Derrick's
theorem \cite{Derrick}. A solution $\phi(r)$ is allowed to be distorted by
making the replacements $\phi(r)\rightarrow\phi_{\lambda}(r)=\phi(\lambda r)$
and $\mathcal{E}\rightarrow\mathcal{E}_{\lambda}$ where $\mathcal{E}_{\lambda
}$ is the energy parameter $\mathcal{E}$ with $\phi$ replaced by
$\phi_{\lambda}$. Upon allowing the parameter $\lambda$ to vary, we require
that a solution representing a stable minimum satisfy $\delta\mathcal{E}=0$
and $\delta^{2}\mathcal{E}\geq0$, or, in terms of $\mathcal{E}_{\lambda}$,
\begin{subequations}
\label{13}%
\begin{align}
\text{(i)\ \ }\frac{d\mathcal{E}_{\lambda}}{d\lambda}\Big|_{\lambda=1}\  &
=0\label{13a}\\
\text{(ii)\ }\frac{d^{2}\mathcal{E}_{\lambda}}{d\lambda^{2}}\Big|_{\lambda=1}
&  \geq0 \label{13b}%
\end{align}

The energy $\mathcal{E}_{\lambda}$ can be written as a sum of two independent
parts, $I_{1\lambda}$, representing the gradient contribution $\mathcal{H}%
_{g}=\frac{1}{2}h(\partial_{r}\phi)^{2}$ to the energy, and $I_{2\lambda}$,
representing the contribution from the potential, $\mathcal{H}_{p}=V(r,\phi)$.
It is shown (see Appendix) that for \textit{any} radially symmetric ansatz
solution satisfying (\ref{10}) with finite energy (or finite energy per unit
length) in a spacetime with a metric of the form given by (\ref{1}), the
stability of the solution, as required by (\ref{13}), is guaranteed.

\bigskip

\ \ Additionally, it is seen that the radial stress vanishes, $T_{r}^{r}=0$,
using the ansatz (\ref{10}). Using $T_{r}^{r}=\partial^{r}\phi\partial_{r}%
\phi-g_{r}^{r}\mathcal{L}$, we find%
\end{subequations}
\begin{equation}
T_{r}^{r}=-\frac{1}{2}h(\partial_{r}\phi)^{2}+V(r,\phi)=0 \label{19}%
\end{equation}

where (\ref{10}) has been used. This indicates a stability against spontaneous
radial collapse or expansion. We also note that the result $T_{rr}=0$ implies%
\begin{equation}
\partial_{r}\phi=\pm\sqrt{2|g_{rr}|V}=\pm(fh)^{-1}X \label{21}%
\end{equation}

(This is a radial generalization of the familiar one-dimensional linear result
$\partial_{x}\phi(x)=\pm\sqrt{2V(\phi)}=\pm\partial_{\phi}W(\phi)$, where
$W(\phi)$ is a superpotential, with $X$ playing the role of $\partial_{\phi}%
W$.) The nonzero stress components for the ansatz solutions are $T_{\xi}^{\xi
}=-g_{\xi}^{\xi}\mathcal{L}=T_{0}^{0}=\mathcal{H}(r)$ (no sum on $\xi$), where
$\xi$ represents a nonradial coordinate, with $T_{\xi}^{\xi}$ independent of
$\xi$.

\bigskip

\ \ We note that for ansatz solutions satisfying (\ref{10}) we have
$\mathcal{H}_{g}=\mathcal{H}_{p}$ and therefore $\mathcal{H}=T_{0}%
^{0}=2V(r,\phi)=\frac{1}{f^{2}(r)h(r)}X^{2}(\phi)$.

\section{Examples}

\ \ Several examples are now given for different metrics, where a form of
$X(\phi)$ is chosen to yield $P(\phi)=\frac{1}{2}X^{2}(\phi)$ corresponding to
the ubiquitous and interesting Higgs-type of potential. The spacetime
background is taken to be fixed - no back reaction of the scalar on the
background geometry is considered. It is assumed that the scalar stress-energy
is negligible in comparison to that of the source, and the scalar field $\phi$
has no direct interaction with the source beyond a response to the background geometry.

\bigskip

\ \ \textbf{(1)}\ \textit{Spherical }$\phi^{4}$\textit{ bubble}: Consider a
spherical scalar field configuration in a flat spacetime with metric given
by\footnote{See Ref.\cite{Bazeia ppt20} (See Section II.B.1) for the example
of a charge immersed in a medium with electric permittivity controlled by real
scalar field.}%
\begin{equation}
ds^{2}=dt^{2}-dr^{2}-r^{2}(d\theta^{2}+\sin^{2}\theta d\varphi^{2})\label{22}%
\end{equation}

In this case $f(r)=r^{2}=\sqrt{g}/\sin\theta$ and $h=|g^{rr}|=1$. We choose%

\begin{equation}
X(\phi)=\lambda(\eta^{2}-\phi^{2}),\ \ \ \ V(r,\phi)=\frac{1}{2f^{2}h}%
X^{2}(\phi)=\frac{\lambda^{2}}{2r^{4}}(\eta^{2}-\phi^{2})^{2} \label{23}%
\end{equation}

where (\ref{9}) has been used, and we take $\lambda$ and $\eta$ to be positive
constants. With (\ref{11}), $\int\frac{d\phi}{X(\phi)}=\int\frac{dr}%
{f(r)h(r)}$ gives
\begin{equation}
\int\frac{d\phi}{(\eta^{2}-\phi^{2})}=\lambda\int\frac{dr}{r^{2}}%
\implies-\frac{1}{\eta}\tanh^{-1}(\phi/\eta)=-\lambda(\frac{1}{r}+C)
\label{24}%
\end{equation}

Setting the integration constant $C=-1/R$ gives the solution%
\begin{equation}
\phi(r)=\eta\tanh\Big[k\Big(\frac{1}{r}-\frac{1}{R}\Big)\Big],\ \ \ k=\lambda
\eta\equiv r_{0} \label{25}%
\end{equation}

With this solution we have $\phi(r)$ remaining everywhere finite. For
$\phi(r)=+\eta\tanh[k(\frac{1}{r}-\frac{1}{R})]$ we have
\begin{equation}
\phi(r)\rightarrow\left\{
\begin{array}
[c]{ll}%
+\eta, & r\rightarrow0\\
-\eta\tanh(\frac{k}{R}), & r\rightarrow\infty
\end{array}
\right\}  \label{26}%
\end{equation}
The solution $\phi/\eta$ is a monotonically decreasing function of $r$ with
asymptotic value of $\phi\rightarrow-\eta\tanh(\frac{k}{R})$ (Fig1). For
$\phi$ and $\eta$ having canonical mass dimension 1, we have a mass dimension
of $-1$ for $k=\lambda\eta$, so that $\lambda$ has mass dimension $-2.$ We can
write $\lambda\eta=k=r_{0}$ where $r_{0}$ is some radial constant. The
configuration (\ref{25}) suggests the existence of a bubble wall centered
somewhere near $r\sim R$ where the energy density $\mathcal{H}(r)$ maximizes
(Fig1). This is a static, radially stable solution to the equation of motion.

\bigskip

\ \ The energy density (\ref{12}) of the solution (\ref{25}) is%
\begin{equation}
\mathcal{H}=\frac{1}{r^{4}}X^{2}(\phi)=\frac{\lambda^{2}}{r^{4}}(\eta^{2}%
-\phi^{2})^{2}=\frac{k^{2}\eta^{2}}{r^{4}}\text{sech}^{4}[k(\frac{1}{r}%
-\frac{1}{R})] \label{27}%
\end{equation}

This maximizes at a finite value of $r\lesssim R$, so that a bubble wall
appears near this radius (Fig1).

\begin{figure}[tbh]
\centering
\includegraphics[width=8.0cm]{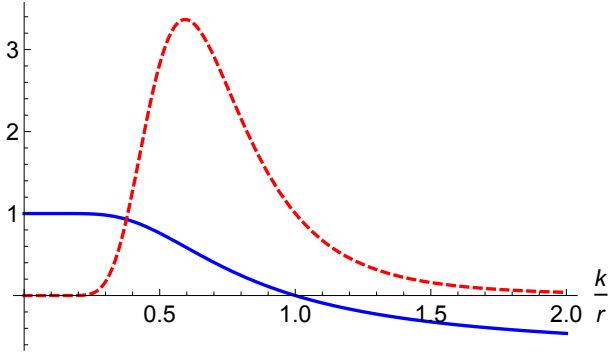}\caption{$\phi(r)/\eta$ (solid) and
$(k/\eta)^{2}\mathcal{H}(r)$ (dashed) vs $k/r$ with $k/R=1$.}%
\end{figure}

\ \ The total configuration energy (mass), $\mathcal{M}=\int d^{3}x\sqrt
{g}\mathcal{H}=4\pi\mathcal{E}$, is given by%
\begin{align}
\mathcal{M}(R)  &  =4\pi\lambda^{2}\int dr\frac{1}{r^{2}}(\eta^{2}-\phi
^{2})^{2}=4\pi\lambda^{2}\eta^{4}\int_{0}^{\infty}dr\frac{1}{r^{2}}%
\text{sech}^{4}\Big(\frac{k}{r}-\frac{k}{R}\Big)\nonumber\\
&  =\frac{4\pi k\eta^{2}}{3}\Big[2+3\tanh\Big(\frac{k}{R}\Big)-\tanh
^{3}\Big(\frac{k}{R}\Big)\Big] \label{28}%
\end{align}

The total mass of the bubble $\mathcal{M}(R)$ decreases monotonically with
$R$. This might lead one to assume that the bubble would tend to expand
radially to decrease its mass, but the stability arguments, including the fact
that $T_{rr}=0$, indicate otherwise. A bubble that is initially formed with a
radius $R$ maintains that radius, and larger bubbles will be less massive.
This is opposite to the case of a spherical bubble formed from a standard
domain wall with a canonical potential $U(\phi)\sim g^{2}(\eta^{2}-\phi
^{2})^{2}$ with no explicit $r$ dependence. In the standard type of scenario
the bubble experiences a radially inward force due to the surface tension,
causing it to collapse. Therefore, without some stabilizing mechanism, the
solution $\phi=\phi(r,t)$ must be time dependent.

\bigskip

\ \ It can be noted that these results essentially duplicate those found in
Ref.\cite{Bazeia PRD18} for the source field $\phi(r)$ of a magnetic monopole
with internal structure\footnote{See Section 3A of Ref. \cite{Bazeia PRD18}.
There, the choice of $R=0$ is made and parameters have been rescaled, and
$X(\phi)$ corresponds to $W_{\phi}(\phi)$.}. That is, the spherical shell
$\phi(r)$ serves as the source field of a monopole for the model in
\cite{Bazeia PRD18}. Additionally, these results basically reproduce those
obtained for the scalar field in Ref.\cite{Bazeia ppt20} regarding
electrically charged solitonic structures\footnote{See Section II.B.1 of
Ref.\cite{Bazeia ppt20}. There, again, the choice of $R=0$ is made and
parameters have been rescaled, and $X(\phi)$ corresponds to $W_{\phi}$.}.

\bigskip

\ \ \textbf{(2)}\ \textit{Schwarzschild }$\phi^{4}$\textit{ bubble}%
:\ \ Consider a spherical scalar field configuration centered on a black hole
with a Schwarzschild radius $r_{S}$ in a Schwarzschild spacetime with metric
described by%
\begin{equation}
ds^{2}=\left(  1-\frac{r_{S}}{r}\right)  dt^{2}-\left(  1-\frac{r_{S}}%
{r}\right)  ^{-1}dr^{2}-r^{2}(d\theta^{2}+\sin^{2}\theta d\varphi^{2})
\label{29}%
\end{equation}

In this case we have%
\begin{equation}
f(r)=r^{2},\ \ \ h(r)=A(r)=\left(  1-\frac{r_{S}}{r}\right)  ,\ \ \ \ f^{2}%
(r)h(r)=r^{4}A=r^{4}\left(  1-\frac{r_{S}}{r}\right)  \label{30}%
\end{equation}

Again, let's choose a $\phi^{4}$ potential with%
\begin{equation}
X(\phi)=-\lambda(\eta^{2}-\phi^{2}),\ \ \ \ V(r,\phi)=\frac{1}{2f^{2}%
(r)h(r)}X^{2}(\phi)=\frac{\lambda^{2}A^{-1}(r)}{2r^{4}}(\eta^{2}-\phi^{2})^{2}
\label{31}%
\end{equation}

With (\ref{11}), $\int\frac{d\phi}{X(\phi)}=\int\frac{dr}{f(r)h(r)}$ gives
$\int\frac{d\phi}{(\eta^{2}-\phi^{2})}=\lambda\int\frac{A^{-1}(r)\ }{r^{2}}%
dr$, so that,\ \ \ \ \ \ \ \ \ \ \ \ \ \ \ \ \ \ \ \ \
\begin{equation}
\frac{\phi(r)}{\eta}=\psi(r)=\tanh\left[  K\ln A(r)\right]  ,\ \ \ (K=\frac
{\lambda\eta}{r_{S}}) \label{32}%
\end{equation}

where the integration constant has been set to zero in this case and
$K\equiv\lambda\eta/r_{S}$. The function $\psi(r)$ is a finite, bounded
function of $r$, with $\psi(r)$ defined for $r\in(r_{S},\infty)$, with
$\psi\rightarrow0$ as $r\rightarrow\infty$ (Fig2). The energy density
$\mathcal{H}=T_{0}^{0}$ is $\mathcal{H}(r,\phi)=\frac{A^{-1}(r)}{f^{2}%
(r)}X^{2}(\phi)=\frac{A^{-1}}{r^{4}}X^{2}$, i.e.,%

\begin{equation}
\mathcal{H}=\frac{A^{-1}}{r^{4}}X^{2}(r)=A^{-1}r^{-4}B(1-\psi^{2}%
)^{2},\ \ \ (B=\lambda^{2}\eta^{4}) \label{34}%
\end{equation}

The energy density is finite for all $r\geq r_{S}$ (i.e., outside the
Schwarzschild horizon), with a maximum beyond $r_{S}$, and $\mathcal{H}%
\rightarrow0$ as $r\rightarrow\infty$ (Fig2).

\begin{figure}[tbh]
\centering
\includegraphics[width=8.0cm]{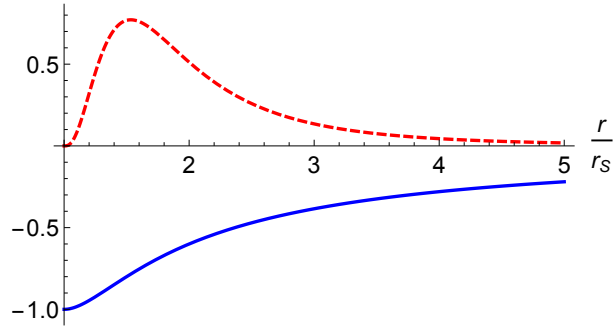}\caption{Sketches of $\phi(r)/\eta$
(solid) and $r_{S}^{4}\mathcal{H}(r)$ (dashed) vs $r/r_{S}$ with $B=10$.}%
\end{figure}

The total energy (mass) of this scalar field configuration is $\mathcal{M}%
=\int_{r_{S}}^{\infty}d^{3}x\sqrt{g}\mathcal{H}(r)$. Using $\sqrt{g}=r^{2}%
\sin\theta$ gives%
\begin{equation}
\mathcal{M}=4\pi\int_{r_{S}}^{\infty}\mathcal{H}(r)\ r^{2}dr=\frac{8\pi
B}{3Kr_{S}}=\frac{8\pi\lambda^{2}\eta^{4}}{3Kr_{S}}=\frac{8\pi\lambda\eta^{3}%
}{3} \label{35}%
\end{equation}

\ \ \textbf{(3)}\ \textit{Cosmic string background}:\ \ Here, the spacetime
metric sourced by a straight cosmic string along the $z$ axis is described by
\cite{Gott ApJ85},\cite{Hiscock PRD85},\cite{Aryal PRD86},\cite{Trenda EJP}%
\begin{equation}
ds^{2}=dt^{2}-dr^{2}-b^{2}r^{2}d\varphi^{2}-dz^{2} \label{36}%
\end{equation}

where $b=(1-4G\mu)$ with $\mu$ being the mass per unit length of the string,
and $\varphi\in\lbrack0,2\pi)$. (One can also define $\varphi^{\prime
}=b\varphi$ with $\varphi^{\prime}\in\lbrack0,2\pi b)$. For $b=1$ we have a
flat spacetime.) In this case%
\begin{equation}
f(r)=\sqrt{g}=br,\ \ h(r)=B^{-1}(r)=A(r)=1 \label{37}%
\end{equation}

We again choose a $\phi^{4}$ potential, and define $\psi(r)=\phi(r)/\eta$,%
\begin{equation}
X(\phi)=-\lambda(\eta^{2}-\phi^{2})=-\lambda\eta^{2}(1-\psi^{2}%
),\ \ \ \ V(r,\phi)=\frac{1}{2f^{2}h}X^{2}=\frac{\lambda^{2}}{2b^{2}r^{2}%
}(\eta^{2}-\phi^{2})^{2} \label{38}%
\end{equation}

From (\ref{11}) we obtain%
\begin{equation}
\psi(r)=\frac{(\rho^{-2\alpha}-1)}{(\rho^{-2\alpha}+1)},\ \ \ \rho\equiv
\frac{r}{r_{0}},\ \ \ (\alpha\equiv\frac{\lambda\eta}{b}) \label{39}%
\end{equation}

where $r_{0}$ is an integration constant. We have $\mathcal{H}=T_{0}%
^{0}=-\mathcal{L}=\frac{1}{2}h(\partial_{r}\phi)^{2}+V=\frac{1}{f^{2}h}%
X^{2}(\phi)$, or
\[
\mathcal{H}=T_{0}^{0}=\frac{1}{b^{2}r^{2}}X^{2}=\frac{\beta}{\rho^{2}}%
(1-\psi^{2})^{2},\ \ \ \beta=\Big(\frac{\lambda^{2}\eta^{2}}{b^{2}}%
\Big)\frac{\eta^{2}}{r_{0}^{2}}=\frac{\alpha^{2}\eta^{2}}{r_{0}^{2}}%
\]
The functions $\psi$ and $\mathcal{H}$ are finite for all $\rho\in
\lbrack0,\infty)$, with $\mathcal{H}$ maximizing at some finite radius with
$\rho=1$, locating a \textquotedblleft wall\textquotedblright\ of the
cylindrical shell (Fig3). The scalar field configuration has a finite
energy/length $\Lambda$ given by%
\begin{equation}
\Lambda=2\pi br_{0}^{2}\int_{0}^{\infty}\mathcal{H}(\rho)\rho d\rho=\frac{8\pi
b\beta r_{0}^{2}}{3\alpha} \label{41}%
\end{equation}

\begin{figure}[tbh]
\centering
\includegraphics[width=8.0cm]{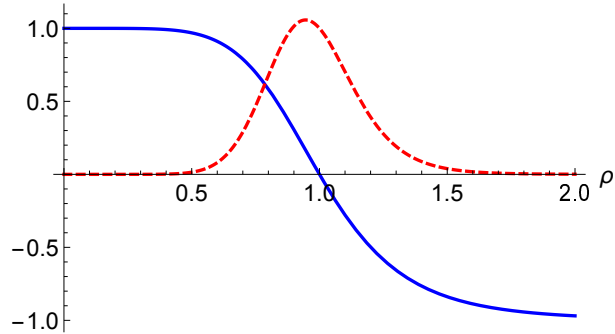}\caption{Sketches of $\phi(\rho
)/\eta$ (solid) and $\mathcal{H}(\rho)/\beta$ (dashed) vs $\rho$.}%
\end{figure}

\ \ Again, we can notice that the results obtained here for a cylindrical
shell in flat spacetime ($b=1$) appear to be in agreement with those found in
Ref \cite{Bazeia PRR19} for the source field $\chi(r)$ for multilayered
vortices\footnote{See Section II.A.1 of Ref .\cite{Bazeia PRR19}.}. This shell
of a neutral scalar field is responsible for the structure of a vortex. The
results reported here are also in apparent agreement with those\footnote{See
Section II.A.1 of Ref. \cite{Bazeia ppt20}.} of Ref. \cite{Bazeia ppt20}.

\bigskip

\ \ \textbf{(4)} \textit{Wormhole background}:\ \ Next, consider a scalar
field $\phi(r)$ in the background spacetime of an
Ellis-Bronnikov-Morris-Thorne wormhole \cite{Ellis},\cite{Bronnikov}%
,\cite{M-T}. The wormhole metric is given by%
\begin{equation}
ds^{2}=dt^{2}-dr^{2}-(r^{2}+a^{2})(d\theta^{2}+\sin^{2}\theta d\varphi^{2})
\label{42}%
\end{equation}

where $r\in(-\infty,\infty)$ and the parameter $a$ represents the
\textquotedblleft radius\textquotedblleft\ of the wormhole throat where $r=0$.
For this case we have $\sqrt{g}=(r^{2}+a^{2})\sin\theta$, with
\begin{equation}
f(r)=(r^{2}+a^{2}),\ \ \ \ A=h=1 \label{43}%
\end{equation}

Again we choose a $\phi^{4}$ potential, with%
\begin{equation}
X=\lambda(\eta^{2}-\phi^{2})=\lambda\eta^{2}(1-\psi^{2});\ \ \ V(r,\phi
)=\frac{1}{2f^{2}h}X^{2}=\frac{\lambda^{2}(\eta^{2}-\phi^{2})^{2}}%
{2(r^{2}+a^{2})^{2}} \label{44}%
\end{equation}

where $\psi(r)=\phi(r)/\eta$. We define the dimensionless radial variable
$\rho\equiv r/a$ and the dimensionless constant $K=\lambda\eta/a$ so that%
\begin{equation}
V(\rho,\psi)=\frac{K^{4}}{2\lambda^{2}}\frac{(1-\psi^{2})^{2}}{(\rho
^{2}+1)^{2}} \label{45}%
\end{equation}

Using (\ref{11}), along with (\ref{43}) and (\ref{44}), then yields%
\begin{equation}
\psi(\rho)=\tanh\left[  K(\tan^{-1}\rho-\tan^{-1}\rho_{0})\right]  \label{46}%
\end{equation}

where $-K\tan^{-1}\rho_{0}$ is an integration constant, with $\rho_{0}%
=r_{0}/a$ representing the center of the scalar field cloud $\psi(\rho)$ where
$\psi(\rho_{0})=0$.

\bigskip

\ \ The energy density is represented by $\mathcal{H}=T_{0}^{0}=\frac{1}%
{2}(\partial_{r}\phi)^{2}+V(r,\phi)$, which for ansatz solutions satisfying
(\ref{10}) becomes $\mathcal{H}=2V(r,\phi)$. From (\ref{45}) and (\ref{46}),%
\begin{equation}
\mathcal{H}=T_{0}^{0}=\frac{K^{4}}{\lambda^{2}}\frac{(1-\psi^{2})^{2}}%
{(\rho^{2}+1)^{2}} \label{47}%
\end{equation}
The functions $\psi(\rho)$ and $\mathcal{H}(\rho)$ are finite for all $\rho$,
with $\mathcal{H}$ maximizing at $\rho_{0}$ where $\psi$ vanishes (Fig4).

\begin{figure}[tbh]
\centering
\includegraphics[width=8.0cm]{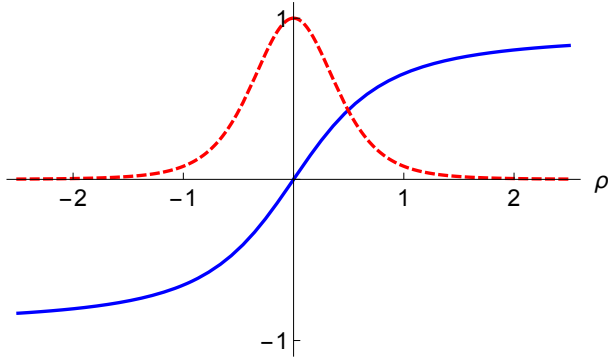}\caption{Sketches of $\psi(\rho)$
(solid) and $(\lambda^{2}/K^{4})\mathcal{H}(\rho)$ (dashed) vs $\rho$. The
center of the cloud is chosen to be centered on the wormhole throat, $\rho
_{0}=0$.}%
\end{figure}

\ \ The mass $M$ of the scalar field configuration occupying the $r\geq0$
region of the spacetime is%
\begin{equation}
M=\int\mathcal{H}(r)\sqrt{g}d^{3}x=4\pi\int_{0}^{\infty}\mathcal{H}%
(r)(r^{2}+a^{2})dr=4\pi a^{3}\int_{0}^{\infty}\mathcal{H}(\rho)(\rho
^{2}+1)d\rho\label{48}%
\end{equation}

Using (\ref{46}) and (\ref{47}) then gives the result%
\begin{equation}
M=\frac{4\pi a^{3}K^{3}}{3\lambda^{2}}\left[  2+\text{sech}^{2}\frac{K\pi}%
{2}\tanh\frac{K\pi}{2}\right]  \label{49}%
\end{equation}

\bigskip

\section{Summary}

\ \ A set of scalar field potentials having the noncanonical form
$U(r,\phi)=r^{-N}P(\phi)$, where $N\in\mathbb{Z}^{+}$ is a positive integer,
and $P(\phi)=\frac{1}{2}W_{\phi}^{2}(\phi)$ with $W(\phi)$ a superpotential,
was introduced in Ref. \cite{Bazeia PRL03}. Physical motivations include
possible descriptions of effective potentials arising from other fields
interacting with the scalar field $\phi$. Additionally, potentials of this
form have been imposed in scalar field theories in order to produce field
theoretic models with new and different features. This has proven to be of
value in subsequent investigations of various models. (See, for example,
\cite{Bazeia PRD18}-\cite{Bazeia ppt20}.) The function $U(r,\phi)$ is
applicable to situations where there is a spherical symmetry in $D$ space
dimensions of flat spacetimes, provided that $N$ and $D$ satisfy certain constraints.

\bigskip

\ Here, a new set of effective potentials is introduced, taking the general
form $V(r,\phi)=F(r)P(\phi)$ where $F(r)=(f^{2}(r)h(r))^{-1}$ is a function of
a radial coordinate $r$ (i.e., spherical or cylindrical symmetry) in a four
dimensional spacetime, with $f(r)$ and $h(r)$ determined by the spacetime
metric $g_{\mu\nu}$. (Specifically, $f(r)$ is the radial part of $\sqrt{g}$
and $h(r)=|g^{rr}|$.) The $\phi$ dependent part of the potential is given by
$P(\phi)=\frac{1}{2}X^{2}(\phi)$, where $X(\phi)$ is a function taking the
role of $W_{\phi}$. This form of potential $V(r,\phi)$ coincides with the
potential $U(r,\phi)$ in the case of a flat four dimensional spacetime with
spherical symmetry. Thus, at least in the case of four dimensions, the set of
potentials $V(r,\phi)$ includes and generalizes the set of potentials
$U(r,\phi)$ of \cite{Bazeia PRL03}. Both types of potentials are of interest
from both physical and mathematical points of view, allowing stable, energy
minimizing radial solutions derivable from a first order differential equation.

\bigskip

\ \ The utility of incorporating a potential $V(r,\phi)$ in a scalar field
theory can be illustrated by using the method introduced by Atmaja and
Ramadhan \cite{Atmaja PRD14} whereby the second order Euler-Lagrange equation
of motion for $\phi(r)$ can be reduced to a first order Bogomol'nyi equation
yielding a BPS type of minimal energy solution for the potential. Moreover, in
Section 4 and the appendix, a general expression for the energy has been
obtained, along with a proof, along the lines of Derrick's theorem
\cite{Derrick}, that the solution $\phi(r)$ is radially stable. Examples of
applying this method with potentials $V(r,\phi)$ have been provided in Section
5, which include using the ubiquitous symmetry breaking $\phi^{4}$ potential
(where $X(\phi)=\lambda(\eta^{2}-\phi^{2})$) in background spacetimes (flat,
Schwarzschild, cosmic string, and wormhole) with radial symmetry (spherical or
cylindrical). The results of examples (1) and (3) presented here are seen to
coincide with those of \cite{Bazeia PRD18} and \cite{Bazeia PRR19} for models
describing magnetic monopoles with internal structure \cite{Bazeia PRD18} and
multilayered vortices \cite{Bazeia PRR19} and charged solitons \cite{Bazeia
ppt20}.

\appendix{}

\section{Stability considerations}

\ \ We consider 4D metrics with radial symmetry (spherical or cylindrical) of
the form%
\begin{equation}
\left\{
\begin{array}
[c]{ll}%
ds^{2}=A(r)dt^{2}-B(r)dr^{2}-R^{2}(r)(d\theta^{2}+\sin^{2}\theta d\varphi
^{2}), & \text{(spherical symmetry) }\medskip\\
ds^{2}=A(r)dt^{2}-B(r)dr^{2}-\rho^{2}(r)d\varphi^{2}-\zeta^{2}(r)dz^{2}, &
\text{(cylindrical symmetry)}%
\end{array}
\right.  \label{A1}%
\end{equation}

in which case we have $\sqrt{g}=\sqrt{AB}R^{2}\sin\theta$ for spherical
symmetry and $\sqrt{g}=\sqrt{AB}\rho\zeta$ for cylindrical symmetry
\cite{Trenda EJP}. (The functions $A(r)$ and $B(r)$ for the cylindrical case
are generally different from those for the spherical case.) We represent the
radial part of $\sqrt{g}$ by%
\begin{equation}
f(r)=\sqrt{A(r)B(r)C(r)} \label{A2}%
\end{equation}

where $\sqrt{C(r)}$ is given by%
\begin{equation}
\sqrt{C(r)}=\left\{
\begin{array}
[c]{ll}%
R^{2}(r), & \text{(spherical symmetry)\medskip}\\
\rho(r)\zeta(r), & \text{(cylindrical symmetry)}%
\end{array}
\right.  \label{A3}%
\end{equation}

Our ansatz for radially symmetric solutions is given by
\begin{subequations}
\label{A4}%
\begin{align}
\partial_{r}\phi &  =\frac{1}{f(r)h(r)}X(\phi),\label{A4a}\\
V(r,\phi)  &  =\frac{1}{f^{2}(r)h(r)}P(\phi)=\frac{1}{2f^{2}(r)h(r)}X^{2}%
(\phi) \label{A4b}%
\end{align}

where $h(r)\equiv|g^{rr}|=B^{-1}(r)$, and $P(\phi)=\frac{1}{2}X^{2}(\phi)$.
The function $X(\phi)$ must be chosen to yield a finite energy (or finite
energy per unit length) solution.

\bigskip

\ The energy of a spherically symmetric solution is%
\end{subequations}
\begin{equation}
E=\int T_{0}^{0}\sqrt{g}d^{3}x=\Omega\int T_{0}^{0}f(r)dr \label{A5}%
\end{equation}

where $\Omega=\int d\Omega=\int\int\sin\theta d\theta d\varphi$. (The solid
angle factor $\Omega$ takes a value of $4\pi$ in a flat spacetime, but may
differ from $4\pi$ in spacetimes with a solid angular deficit or surplus.)

\bigskip

\ In the case of cylindrical symmetry, we can define the energy in a length
$L$ along the $z$ direction as%
\begin{equation}
E=\omega L\int T_{0}^{0}f(r)dr \label{A6}%
\end{equation}

where $\omega=\int d\varphi$. (We have $\omega=2\pi$ for a flat spacetime with
no angular deficit or surplus.) For either case, let us define the quantity%
\begin{equation}
\mathcal{E}=\left\{
\begin{array}
[c]{ll}%
\dfrac{E}{\Omega},\smallskip & \text{(spherical symmetry)}\\
& \text{or}\\
\dfrac{E}{\omega L}, & \text{(cylindrical symmetry)}%
\end{array}
\right.  \label{A7}%
\end{equation}

so that, in either case,%
\begin{equation}
\mathcal{E}=\int T_{0}^{0}f(r)dr=\int\mathcal{H}f(r)dr \label{A8}%
\end{equation}

with $\mathcal{H}\equiv T_{0}^{0}$.

\bigskip

\ \ To investigate solution stability, we demand that $\mathcal{E}$ be finite,
and that, furthermore, the solution considered represents a stable minimum for
the energy. We follow the procedure used in Derrick's theorem \cite{Derrick}
requiring that $\delta\mathcal{E}=0$ and $\delta^{2}\mathcal{E}\geq0$ for a
stable static solution that minimizes the action. To do so, we define
$\phi_{\lambda}(r)=\phi(\lambda r)=\phi(r^{\prime})$, where $r^{\prime
}=\lambda r$ with $\lambda$ being an arbitrary real parameter. We then define
the energy parameter $\mathcal{E}_{\lambda}$ with $\phi(r)\rightarrow
\phi_{\lambda}(r)$ in the energy integral. For stability, we require
\begin{subequations}
\label{A9}%
\begin{align}
\text{(i)\ \ }\frac{d\mathcal{E}_{\lambda}}{d\lambda}\Big|_{\lambda=1}\  &
=0\label{A9a}\\
\text{(ii)\ }\frac{d^{2}\mathcal{E}_{\lambda}}{d\lambda^{2}}\Big|_{\lambda=1}
&  \geq0 \label{A9b}%
\end{align}

Now, for \textit{any} static, radially symmetric solution to the second order
equation of motion $\square\phi+\partial_{\phi}V(r,\phi)=0$, for which
$\mathcal{L}=\frac{1}{2}\partial^{r}\phi\partial_{r}\phi-V(r,\phi)$ and
$\mathcal{H}=T_{0}^{0}=-g_{0}^{0}\mathcal{L}=-\mathcal{L}$, i.e.,
$\mathcal{H}=\frac{1}{2}h(\partial_{r}\phi)^{2}+V(r,\phi)$, where $V(r,\phi)$
is given by (\ref{A4}), the energy integral can be written as a sum of
gradient plus potential contributions, $\mathcal{E}=I_{1}+I_{2}$:%
\end{subequations}
\begin{equation}
\mathcal{E}=I_{1}+I_{2};\ \ \ \ \ I_{1}=\int\frac{1}{2}(\partial_{r}\phi
)^{2}G(r)dr,\ \ \ I_{2}=\int P(\phi)H(r)dr \label{A10}%
\end{equation}

where we define $G=hf$, and $H=1/(fh)=G^{-1}$:
\begin{equation}
G(r)=hf=\sqrt{\frac{AC}{B}},\ \ \ H(r)=\frac{1}{hf}=\sqrt{\frac{B}{AC}}%
=G^{-1}(r) \label{A11}%
\end{equation}

Upon making the replacement $\phi\rightarrow\phi_{\lambda}$ we have
$\mathcal{E}\rightarrow\mathcal{E}_{\lambda}=I_{1\lambda}+I_{2\lambda}$, with%
\begin{equation}%
\begin{array}
[c]{ll}%
I_{1\lambda} & =\int\frac{1}{2}(\partial_{r}\phi_{\lambda})^{2}G(r)dr=\lambda
\int\frac{1}{2}(\partial_{r^{\prime}}\phi_{\lambda})^{2}G(r)dr^{\prime}%
\equiv\lambda J_{1}(r)\\
I_{2\lambda} & =\int P(\phi_{\lambda})H(r)dr=\lambda^{-1}\int P(\phi_{\lambda
})H(r)dr^{\prime}\equiv\lambda^{-1}J_{2}(r)
\end{array}
\label{A12}%
\end{equation}

The integrals $J_{1}$ and $J_{2}$ are functions of $r=\lambda^{-1}r^{\prime}$.
Therefore, derivatives $\partial_{\lambda}J_{1,2}(r)=$ $\partial_{\lambda
}J_{1,2}(\lambda^{-1}r^{\prime})$ involve $\partial_{\lambda}G(r)$ and
$\partial_{\lambda}H(r)$, with $r=\lambda^{-1}r^{\prime}$, where%
\begin{equation}
\frac{dG(r)}{d\lambda}=\frac{dG(r)}{dr}\frac{dr}{d\lambda}=-\lambda
^{-2}r^{\prime}G^{\prime}(r),\ \ \ \frac{dH(r)}{d\lambda}=-\lambda
^{-2}r^{\prime}H^{\prime}(r) \label{A13}%
\end{equation}

where we denote $G^{\prime}(r)=\partial_{r}G(r)$, $H^{\prime}(r)=\partial
_{r}H(r)$.

\bigskip

\ \ Now using $\mathcal{E}_{\lambda}=I_{1\lambda}+I_{2\lambda}$ along with
some straightforward (but a little tedious) algebra, we arrive at
\begin{subequations}
\label{A14}%
\begin{align}
\frac{d\mathcal{E}_{\lambda}}{d\lambda}  &  =J_{1}-\lambda^{-1}K_{1}%
-\lambda^{-2}J_{2}-\lambda^{-3}Q_{1}\label{A14a}\\
\frac{d^{2}\mathcal{E}_{\lambda}}{d\lambda^{2}}  &  =\lambda^{-3}%
K_{2}+2\lambda^{-3}J_{2}+4\lambda^{-4}Q_{1}+\lambda^{-5}Q_{2} \label{A14b}%
\end{align}

where%
\end{subequations}
\begin{equation}%
\begin{array}
[c]{ll}%
J_{1}=\int\frac{1}{2}(\partial_{r^{\prime}}\phi_{\lambda})^{2}G(r)dr^{\prime
}, & J_{2}=\int P(\phi_{\lambda})H(r)dr^{\prime}\\
K_{1}=\int\frac{1}{2}(\partial_{r^{\prime}}\phi_{\lambda})^{2}r^{\prime
}G^{\prime}(r)dr^{\prime}, & Q_{1}=\int P(\phi_{\lambda})r^{\prime}H^{\prime
}(r)dr^{\prime}\\
K_{2}=\int\frac{1}{2}(\partial_{r^{\prime}}\phi_{\lambda})^{2}r^{\prime
2}G^{\prime\prime}(r)dr^{\prime}, & Q_{2}=\int P(\phi_{\lambda})r^{\prime
2}H^{\prime\prime}(r)dr^{\prime}%
\end{array}
\label{A15}%
\end{equation}

\bigskip

\ \ The objective is to evaluate (\ref{A14}) at $\lambda=1$ using (\ref{A15})
evaluated at $\lambda=1$ to verify the stability conditions (\ref{A9}) for all
ansatz solutions satisfying (\ref{A4}), subject to the metric conditions of
(\ref{A1}) - (\ref{A3}). For ansatz solutions satisfying the Bogomol'nyi
equation (\ref{A4a}) with the potential (\ref{A4b}) it is found that
$I_{1}=I_{2}$, i.e., the gradient and potential contributions to
$\mathcal{E}=I_{1}+I_{2}$ are equal. Furthermore, setting $\lambda=1$ for the
integrals of (\ref{A15}) simply amounts to setting $r^{\prime}=r$. In that
case, it turns out to be convenient to re-express the integrals $J_{1}$,
$K_{1}$, and $K_{2}$, when evaluated at $\lambda=1$, in terms of $P(\phi)$ and
the function $G(r)$. Specifically, using $\frac{1}{2}(\partial_{r}\phi
)^{2}G(r)=P(\phi)H(r)$%
\begin{equation}%
\begin{array}
[c]{ll}%
J_{1}\rightarrow\int(PH^{2})Gdr, & J_{2}=\int PHdr\\
K_{1}\rightarrow\int(PH^{2})rG^{\prime}dr, & Q_{1}=\int P\ rH^{\prime}dr\\
K_{2}\rightarrow\int(PH^{2})r^{2}G^{\prime\prime}dr, & Q_{2}=\int
P\ r^{2}H^{\prime\prime}(r)dr
\end{array}
\label{A16}%
\end{equation}

where $P=P(\phi)$, $G=G(r)$, $H=H(r)$, etc.

\bigskip

\ \ Now evaluating (\ref{A14a}) at $\lambda=1$ and enforcing (\ref{A9a}) gives%
\begin{align}
\frac{d\mathcal{E}}{d\lambda}\Big|_{\lambda=1}  &  =(J_{1}-K_{1}-J_{2}%
-Q_{1})|_{\lambda=1}=(I_{1}-I_{2})|_{\lambda=1}-(K_{1}+Q_{1})|_{\lambda
=1}=0\nonumber\\
&  \implies(K_{1}+Q_{1})|_{\lambda=1}=0 \label{A17}%
\end{align}

In fact, using
\begin{equation}
H=G^{-1},\ \ H^{\prime}=-G^{-2}G^{\prime},\ \ \ H^{\prime\prime}%
=2G^{-3}G^{\prime2}-G^{-2}G^{\prime\prime} \label{A18}%
\end{equation}

one can see that $K_{1}+Q_{1}=\int P\ r(H^{2}G^{\prime}+H^{\prime})dr=0$,
i.e., $\dfrac{d\mathcal{E}}{d\lambda}\Big|_{\lambda=1}$ vanishes identically.

\bigskip

\ \ Next, an evaluation of (\ref{A14b}) at $\lambda=1$ gives%
\begin{equation}
\frac{d^{2}\mathcal{E}_{\lambda}}{d\lambda^{2}}\Big|_{\lambda=1}=(K_{2}%
+2J_{2}+4Q_{1}+Q_{2})|_{\lambda=1} \label{A19}%
\end{equation}

Using the integrals in (\ref{A16}) produces%
\begin{equation}
\frac{d^{2}\mathcal{E}_{\lambda}}{d\lambda^{2}}\Big|_{\lambda=1}=\int
P(\phi)\left[  (r^{2}H^{2}G^{\prime\prime})+(2H)+(4rH^{\prime})+(r^{2}%
H^{\prime\prime})\right]  dr \label{A20}%
\end{equation}

Upon using (\ref{A18}) with a little algebra, this can be reduced to%
\begin{equation}
\frac{d^{2}\mathcal{E}_{\lambda}}{d\lambda^{2}}\Big|_{\lambda=1}=\int
P(\phi)G^{-3}\beta(r)dr \label{A21}%
\end{equation}

where%
\begin{equation}
\beta(r)=r^{2}G^{\prime2}-2rGG^{\prime}+G^{2}=(rG^{\prime}-G)^{2}\geq0
\label{A22}%
\end{equation}

Since $P(\phi)G^{-3}(r)\beta(r)\geq0$ for any ansatz solution, we have the
condition of (\ref{A9b}) being automatically satisfied, $\partial_{\lambda
}^{2}\mathcal{E}_{\lambda}|_{\lambda=1}\geq0$. We then conclude that for
\textit{any} radially symmetric ansatz solution with finite energy (or finite
energy per unit length) in a spacetime with a metric of the form given by
(\ref{A1}), the radial stability of the solution (i.e., stability against
spontaneous radial expansion or collapse), as required by (\ref{A9}), is guaranteed.

\bigskip

\ \ In addition, we can take notice of the vanishing of the radial tension
$T_{r}^{r}$ for the ansatz solutions for both the spherical and cylindrical
symmetries:%
\begin{equation}
T_{r}^{r}=\partial^{r}\phi\partial_{r}\phi-g_{r}^{r}\mathcal{L=-}\frac{1}%
{2}h(\partial_{r}\phi)^{2}+V(r,\phi)=0 \label{A23}%
\end{equation}

This also indicates a stability against spontaneous radial expansion or contraction.


\begin{thebibliography}{99}                                                                                               %


\bibitem {Bazeia PRL03}D. Bazeia, J. Menezes, R. Menezes, \textquotedblleft
New global defect structures\textquotedblright, Phys. Rev. Lett. 91 (2003)
241601 \textbullet\ e-Print: hep-th/0305234 [hep-th]

\bibitem {Derrick}G.H. Derrick, \textquotedblleft Comments on nonlinear wave
equations as models for elementary particles\textquotedblright, J. Math. Phys.
5 (1964) 1252-1254

\bibitem {Bazeia PRD18}D. Bazeia, M. A. Marques, R. Menezes, \textquotedblleft
Magnetic monopoles with internal structure\textquotedblright, Phys. Rev. D 97
(2018) 10, 105024 \textbullet\ e-Print: 1805.03250 [hep-th]

\bibitem {Bazeia PRR19}D. Bazeia, M. A. Liao, M. A. Marques, R. Menezes,
\textquotedblleft Multilayered Vortices\textquotedblright, Phys. Rev. Research
1, 033053 (2019), \textbullet\ e-Print: 1908.07871 [hep-th]

\bibitem {Andrade PRD19}J. Andrade, R. Casana, E. da Hora, C. dos Santos,
\textquotedblleft First-order solitons with internal structures in an extended
Maxwell-CP(2) model\textquotedblright, Phys. Rev. D 99 (2019) 5, 056014
\textbullet\ e-Print: 1901.05094 [hep-th]

\bibitem {Casana PRD20}R. Casana, A. C. Santos, M. L. Dias, \textquotedblleft
BPS solitons with internal structure in the gauged O(3) sigma
model\textquotedblright, Phys. Rev. D 102 (2020) 8, 085002 \textbullet
\ e-Print: 2006.16466 [hep-th]

\bibitem {Bazeia ppt20}D. Bazeia, M. A. Marques, R. Menezes, \textquotedblleft
Electrically charged localized structures\textquotedblright, Eur. Phys. J. C
81 (2021) 1, 94 \textbullet\ e-Print: 2011.01766 [physics.gen-ph]

\bibitem {Gonzalez RMF01}J. A. Gonzalez, D. Sudarsky, \textquotedblleft Scalar
solitons in a four-dimensional curved space-time\textquotedblright, Rev. Mex.
Fis. 47 (2001) 231-233 \textbullet\ e-Print: gr-qc/0102061 [gr-qc]

\bibitem {Perivol PRD18}L. Perivolaropoulos, \textquotedblleft Gravitational
Interactions of Finite Thickness Global Topological Defects with Black
Holes\textquotedblright, Phys. Rev. D 97 (2018) 12, 124035 \textbullet
\ e-Print: 1804.08098 [gr-qc]

\bibitem {Perivol PRD19}G. Alestas, L. Perivolaropoulos, \textquotedblleft
Evading Derrick's theorem in curved space: Static metastable spherical domain
wall\textquotedblright, Phys. Rev. D 99 (2019) 6, 064026 \textbullet\ e-Print:
1901.06659 [gr-qc]

\bibitem {Carloni PRD20}S. Carloni, J. L. Rosa, \textquotedblleft Derrick's
theorem in curved spacetime\textquotedblright, Phys. Rev. D 100 (2019) 2,
025014 \textbullet\ e-Print: 1906.00702 [gr-qc]

\bibitem {Hartmann PRD20}B. Hartmann, G. Luchini, C. P. Constantinidis, C. F.
S. Pereira, \textquotedblleft Real scalar field kinks and antikinks and their
perturbation spectra in a closed universe\textquotedblright, Phys. Rev. D 101
(2020) 7, 076004 \textbullet\ e-Print: 1908.09684 [hep-th]

\bibitem {Perivol PRD20}G. Alestas, G.V. Kraniotis, L. Perivolaropoulos,
\textquotedblleft Existence and stability of static spherical fluid shells in
a Schwarzschild-Rindler--anti--de Sitter metric\textquotedblright, Phys. Rev.
D 102 (2020) 10, 104015 \textbullet\ e-Print: 2005.11702 [gr-qc]

\bibitem {Atmaja PRD14}A. N. Atmaja, H. S. Ramadhan, \textquotedblleft
Bogomol'nyi equations of classical solutions\textquotedblright, Phys. Rev. D
90 (2014) 10, 105009 \textbullet\ e-Print: 1406.6180 [hep-th]

\bibitem {Adam JHEP16}C. Adam, F. Santamaria, \textquotedblleft The
First-Order Euler-Lagrange equations and some of their uses\textquotedblright,
JHEP 12 (2016) 047 \textbullet\ e-Print: 1609.02154 [hep-th]

\bibitem {Atmaja JHEP16}A. N. Atmaja, H. S. Ramadhan, E. da Hora,
\textquotedblleft More on Bogomol'nyi equations of three-dimensional
generalized Maxwell-Higgs model using on-shell method\textquotedblright, JHEP
02 (2016) 117 \textbullet\ e-Print: 1505.01241 [hep-th]

\bibitem {Damour PRL93}T. Damour, G. Esposito-Farese, \textquotedblleft
Nonperturbative strong field effects in tensor - scalar theories of
gravitation\textquotedblright, Phys.Rev.Lett. 70 (1993) 2220-2223

\bibitem {Herdeiro PRD21}C.A.R. Herdeiro, T. Ikeda, M. Minamitsuji, T.
Nakamura, E. Radu, \textquotedblleft Spontaneous scalarization of a conducting
sphere in Maxwell-scalar models\textquotedblright, Phys.Rev.D 103 (2021) 4,
044019 \textbullet\ e-Print: 2009.06971 [gr-qc]

\bibitem {Trenda EJP}C. S. Trendafilova, S. A. Fulling, \textquotedblleft
Static solutions of Einstein's equations with cylindrical
symmetry\textquotedblright, Eur. J. Phys. 32 (2011) 1663-1677 \textbullet
\ e-Print: 1101.4668 [gr-qc]

\bibitem {Gott ApJ85}J. R. Gott, III, \textquotedblleft Gravitational lensing
effects of vacuum strings: Exact solutions\textquotedblright, Astrophys. J.
288 (1985) 422-427

\bibitem {Hiscock PRD85}W. A. Hiscock, \textquotedblleft Exact Gravitational
Field of a String\textquotedblright, Phys. Rev. D 31 (1985) 3288-3290

\bibitem {Aryal PRD86}M. Aryal, L. H. Ford, A. Vilenkin, \textquotedblleft
Cosmic Strings and Black Holes\textquotedblright, Phys. Rev. D 34 (1986) 2263

\bibitem {Ellis}H. G. Ellis, \textquotedblleft Ether flow through a drainhole
- a particle model in general relativity\textquotedblright, J. Math. Phys. 14
(1973) 104-118

\bibitem {Bronnikov}K. A. Bronnikov, \textquotedblleft Scalar-tensor theory
and scalar charge\textquotedblright, Acta Phys. Polon. B 4 (1973) 251-266

\bibitem {M-T}M. S. Morris, K. S. Thorne, \textquotedblleft Wormholes in
space-time and their use for interstellar travel: A tool for teaching general
relativity\textquotedblright, Am. J. Phys. 56 (1988) 395-412
\end{thebibliography}
\end{document}